\documentclass[12pt]{article}
\usepackage[usenames,dvipsnames]{color}
\usepackage{a4wide}
\usepackage{amsmath}
\usepackage{amsfonts}
\usepackage{amssymb}
\usepackage{amsthm}
\usepackage{amsmath}
\usepackage{amsthm}
\usepackage{hyperref}
\usepackage{latexsym}
\usepackage[T1]{fontenc}
\usepackage{color}
\usepackage[titletoc]{appendix}

\newcommand{\eq}[1]{\begin{equation}
                     \begin{split} #1 \end{split}
                     \end{equation}}

\newcommand{\bfrp}{\bar{\mathfrak{p}}^*}

\newcommand{\pxi}{\frac{\partial}{\partial x^i}}
\newcommand{\pxj}{\frac{\partial}{\partial x^j}}
\newcommand{\pxsi}{\frac{\partial}{\partial x^*_i}}
\newcommand{\pxsj}{\frac{\partial}{\partial x^*_j}}
\newcommand{\pxii}{\frac{\partial}{\partial \xi^i}}
\newcommand{\pxij}{\frac{\partial}{\partial \xi^j}}
\newcommand{\pxisi}{\frac{\partial}{\partial \xi^*_i}}
\newcommand{\pxisj}{\frac{\partial}{\partial \xi^*_j}}
\newcommand{\ptxi}{\frac{\partial}{\partial \tilde x_i}}
\newcommand{\ptxj}{\frac{\partial}{\partial \tilde x_j}}

\newtheorem{definition}{Definition}
\newtheorem{theorem}{Theorem}
\newtheorem{rem}[theorem]{Remark}

\def\DFT{Double Field Theory\ }

\begin{document}
\thispagestyle{empty}
\vspace*{-1.5cm}

\begin{flushright}
  {\small
  ITP-UH-24/14\\
  
  }
\end{flushright}

\vspace{1.5cm}

%%%%%%%%%%%%%%%%%%%%%%%%%%%%%%%%%%%%%%%%%%%%%%%
%%%%%%%%%%%%%%%%%%%%%%%%%%%%%%%%%%%%%%%%%%%%%%%

\begin{center}
{\huge
Star products on graded manifolds \\
and $\alpha '$-corrections to Courant algebroids from string theory
}
\end{center}

%%%%%%%%%%%%%%%%%%%%%%%%%%%%%%%%%%%%%%%%%%%%%%%
%%%%%%%%%%%%%%%%%%%%%%%%%%%%%%%%%%%%%%%%%%%%%%%

\vspace{0.4cm}

\begin{center}
  Andreas Deser 
 
\end{center}

%%%%%%%%%%%%%%%%%%%%%%%%%%%%%%%%%%%%%%%%%%%%%%%
%%%%%%%%%%%%%%%%%%%%%%%%%%%%%%%%%%%%%%%%%%%%%%%

\vspace{0.4cm}

\begin{center} 
\small{\emph{Institut f\"ur Theoretische Physik } and \emph{Riemann Center for Geometry and Physics}, \\ 
   {\em Leibniz Universit\"at Hannover}\\

}
\small{Email: {\tt andreas.deser@itp.uni-hannover.de} }

\vspace{0.25cm}

\end{center} 

\vspace{0.4cm}

%%%%%%%%%%%%%%%%%%%%%%%%%%%%%%%%%%%%%%%%%%%%%%%
%%%%%%%%%%%%%%%%%%%%%%%%%%%%%%%%%%%%%%%%%%%%%%%
\begin{center}
\today
\end{center}

\begin{abstract}
Deformation theory refers to an apparatus in many parts of math and physics for going from an infinitesimal (= first order) deformation to a full deformation, either formal or convergent appropriately. %\cite{several}.
 If the algebra being deformed is that of \emph{observables}, the result is deformation quantization, independent of any realization in terms of Hilbert space operators.
There are very important but rare cases in which a formula for a full deformation is known.  For physics, the most important is the Moyal-Weyl star product formula.

 In this paper, we concentrate on 
deformations of Courant algebroid  structures via star products on graded manifolds. In particular, we construct a graded version of the Moyal-Weyl star product.
Recently, in \DFT (DFT),  deformations of the C-bracket and $O(d,d)$-invariant bilinear form to first order in the closed string sigma model coupling $\alpha '$ were derived by analyzing the transformation properties of the Neveu-Schwarz $B$-field. By choosing a particular Poisson structure on the Drinfel'd double corresponding to the Courant algebroid structure of the generalized tangent bundle, we reproduce these deformations for a specific solution of the strong constraint of DFT as expansion of a graded version of the Moyal-Weyl star product. 

\end{abstract}

%%%%%%%%%%%%%%%%%%%%%%%%%%%%%%%%%%%%%%%%%%%%%%%
%%%%%%%%%%%%%%%%%%%%%%%%%%%%%%%%%%%%%%%%%%%%%%%

\clearpage

%\trd{a few minor changes in English/exposition}
 %but will leave really trivial editorial comments for a later version; also a few structural comments}
\tableofcontents

%%%%%%%%%%%%%%%%%%%%%%%%%%%%%%%%%%%%%%%%%%%%%%%
%%%%%%%%%%%%%%%%%%%%%%%%%%%%%%%%%%%%%%%%%%%%%%%

\section{Introduction}
Non-linear sigma models for closed strings are defined by maps $X: \Sigma \rightarrow M$ from a Riemann surface $\Sigma$ (called the worldsheet) to a target space $M$ equipped with a Riemannian metric $G \in \Gamma(\otimes^2 T^*M)$, a two-form $B$-field $B\in \Gamma(\wedge^2 T^*M)$ and a dilaton $\phi \in {\cal C}^\infty(M)$. \emph{Classical} closed string theory is given if $\Sigma$ has genus zero and string loop corrections are given by worldsheets with higher genus. Taking classical closed string theory, the defining sigma model itself has a perturbative expansion, determined by the parameter $\alpha '$ whose relation to the fundamental length scale $l_s$ of string theory is determined by $\alpha ' = l_s^2$.  

%\trd{although only the expansion to order  $\alpha ' = l_s^2$ is currently known?}

%\trd{? a perturbative expansion is known to exist by general nonsense but not known as explicit formulas?}

%\tla{Yes, indeed. The $\alpha'$ expansion of the effective target space field theory (explained in the next paragraph how to determine it) is currently known up to oder $(\alpha')^4$, brave people calculated loop diagrams to this order. The Courant algebroid structure was known only to zeroth order and from Barton and Olaf on up to first order. }

%\tgr{It would be good to have a sentence or two for comments like those you wrote above - how results were known in various contexts at first order and in special cases up to order $t^4$?}

The renormalization group flow equations of classical string sigma models lead to a set of partial differential equations for the metric, $B$-field and dilaton which, to lowest order in the expansion parameter $\alpha'$, contain Einstein's equations. The target space \emph{effective field theory} is defined by a field theory on $M$ having the renormalization group equations as its Euler-Lagrange equations. Thus, there is also an $\alpha '$-expansion of the classical effective field theory, whose lowest order is given, for example,  by the well-known type-IIA supergravity action \cite{Hull:1987pc, Ketov:2000dy, Kaloper:1997ux}.

%\trd{are there other known examples?}

%\tla{Yes, e.g. heterotic string theory. But I am not an expert so I should not write more in the introduction.}

On the level of the sigma model, in case the target space has isometries, Busher \cite{Buscher:1987sk} showed the existence of a physically equivalent theory by gauging the isometries (and thus introducing an auxiliary field $\tilde X^i$ for every isometry direction) and integrating out appropriate gauge degrees of freedom. The result is a non-linear sigma model on a target space $\tilde M$ which is defined to be the \emph{T-dual} to $M$. The prescription to explicitly calculate the metric and $B$-field on $\tilde M$ is known as ``Buscher rules''. It is shown e.g. in \cite{Giveon:1994fu}  that the latter are given by the action of the group $O(d,d)$ on a generalized metric in case there are $d$ isometries. $O(d,d)$ is the structure group of the generalized tangent bundle \cite{Hitchin:2004ut, Gualtieri:2003dx}, locally given by $TM \oplus T^*M$; it turns out that the Courant algebroid structure of the latter is the appropriate language to describe T-duality on the target space \cite{Hull:2007zu, Grana:2008yw} . 

The attempt to formulate a classical field theory manifestly invariant under the action of $O(d,d)$ leads to the introduction of \emph{double field theory} (DFT) \cite{Siegel:1993th, Hull:2009zb, Hull:2009mi, Hohm:2010jy, Hohm:2010pp, Zwiebach:2011rg}, in which the winding degrees of freedom of a closed string are interpreted as canonically conjugate to a second set of target space coordinates. The $O(d,d)$-invariant action of DFT reduces to the standard effective type IIA supergravity action by applying the strong constraint, which eliminates half of the configuration space coordinates. In addition the action obeys a gauge symmetry governed by the \emph{C-bracket}, an extension of the Courant bracket of generalized geometry in the sense that it also includes the winding degrees of freedom. 

All of these achievements (and many more) are at the level of the effective target space field theory to \emph{lowest order in} $\alpha '$, in the sense of \cite{Hohm:2014eba}. Clearly, to understand classical string theory and in particular T-duality, it is desirable to extend the structures of generalized geometry and DFT to higher orders in $\alpha '$, or phrased differently, to \emph{deform} the Courant algebroid (and C-bracket) structures encountered so far. The first results in this direction were found in DFT, where in \cite{Hohm:2014eba, Hohm:2014xsa} a consistent deformation of the C-bracket and $O(d,d)$-invariant bilinear pairing was given. It is the main goal of our work to propose a way to understand these deformations in terms of a \emph{star product expansion} on an appropriate graded Poisson manifold. 

%Deformation theory refers to an apparatus in many parts or math and physics for going from an infinitesimal (= first order) deformation to a full deformation, either formal or convergent appropriately. If the algebra being deformed is that of \emph{observables}, the result is deformation quantization, independent of any realization in terms of Hilbert space operators \cite{stern:short}.
%There are very important but rare cases in which  formulas for a full deformation are known.  Of these, the most important for physics is the Moyal-Weyl star product formula. In this paper, we concentrate on 
%deformations of Courant algebroid structures via star products on graded manifolds. In particular, we construct a graded version of the Moyal-Weyl star product.

Since the seminal work \cite{Bayen:1977ha, Bayen:1977hb}, deformation theory became popular in physics, where the algebra being deformed is that of \emph{observables}. Independent of the theory of linear operators \cite{stern:short},  it was possible to give a formulation of quantum mechanics equivalent to the one mostly used in physics. The full deformation of the algebra of functions on phase space is given by the Moyal-Weyl star product formula. In the last two decades, the latter product turned out to be realized in the operator product expansion of open string vertex operators in the presence of a Neveu-Schwarz $B$-field \cite{Schomerus:1999ug, Seiberg:1999vs}. Similar structures in closed string theory are in an active study at the moment \cite{Blumenhagen:2010hj, Lust:2010iy, Blumenhagen:2011ph, Bakas:2013jwa, Blumenhagen:2013zpa, Mylonas:2012pg}. In mathematics, star products were further studied on graded manifolds and it was realized that they are intimately connected to the deformation theory of Courant algebroids \cite{2014arXiv1410.3346G}. It is one of the main intensions of this work to show that these kinds of star products play a role in closed string theory and DFT.  
 
%are an active area of research in mathematics. JIM, I NEED YOUR EXPERIENCE HERE! 

The structure of the paper is the following: In the first part of chapter \ref{review},  we review the reformulation of Lie bialgebroids in terms of homological vector fields on graded manifolds due to Roytenberg \cite{Deethesis, Roytenberg:2002nu, roytenberg:structure}, leading to the introduction of the Drinfel'd double of a Lie bialgebroid. This is followed by a brief review of elements of the gauge algebra of DFT, especially the C-bracket in the second part. Using the language of the Drinfel'd double, we review the representation of the C-bracket in terms of Poisson brackets \cite{Deser:2014mxa} in a derived form, in the sense of \cite{KosmannSchwarzbach:2003en}.  

In chapter \ref{deformation}, we first give the definition of star products and the star commutator as well as their generalizations to graded Poisson manifolds. In the first non-trivial order of the deformation parameter, the star commutator gives the Poisson bracket on the underlying manifold. As a consequence, we interpret higher order corrections to the star commutator as deformations of the Poisson bracket. This, together with the representation of the C-bracket introduced in chapter \ref{review} in terms of Poisson brackets on the Drinfel'd double, enables us to propose a way to explicitly determine deformations of the C-bracket. For computational simplicity, we restrict ourselves to a specific solution to the strong constraint, i.e. we calculate a deformation of the resulting Courant algebroid structure. In an analogous manner, we compute deformations to the $O(d,d)$-invariant bilinear form after expressing it in terms of Poisson brackets. It turns out that these deformations coincide with those found in DFT. 

We conclude by giving an outlook on how the results could be extended to more general Poisson structures, leading to the introduction of fluxes. We remark why the name flux is justified by comparing their local expressions to the corresponding objects used in string compactifications.

\emph{Acknowledgements:}  The author wishes to thank Jim Stasheff for input on the classical deformation theory and to thank Dee Roytenberg, Marco Zambon, Barton Zwiebach, Peter Schupp, Andre Coimbra, Olaf Hohm and Erik Plauschinn for discussions. 

\section{Poisson brackets and the gauge algebra of DFT}
\label{review}
In order to apply techniques of deformation theory to objects arising in DFT, a bridge has to be built between structures in DFT such as the C-bracket and concepts of symplectic geometry, e.g. Poisson brackets. By identifying the notion of double field as  a function on the Drinfel'd double of a particularly adapted Lie bialgebroid, this was achieved in \cite{Deser:2014mxa}. In the following, we review results of this work with regard to a formulation suitable for deformation theory. 

\subsection{The Drinfel'd double of a Lie bialgebroid and double fields}
Lie bialgebroids \cite{Mackenzie:GT} and Courant algebroids \cite{zbMATH00004959} are central structures in the generalized geometry and DFT-description of configuration spaces and observables arising in compactifications of string theory.

\subsubsection{Lie algebroids and parity reversal}
 Starting with a vector bundle $A$ over a base manifold $M$, the differential graded algebra $\Gamma(\wedge^\bullet A^*)$ of sections in $\wedge^\bullet A^*$ can be identified with the polynomials on the parity shifted version of $A$ with coefficients depending on the base\footnote{For a $\mathbb{Z}_2$-graded vector space $V = V_0 \oplus V_1$ with even elements in $V_0$ and odd elements in $V_1$, the parity reversion $\Pi$ is defined by 
\eq{
(\Pi V)_0 = V_1 \quad \textrm{and} \quad (\Pi V)_1 = V_0\;. \nonumber 
} 
}:
\eq{
\label{observe}
\Gamma(\wedge^\bullet A^*) \simeq \textrm{Pol}^\bullet(\Pi A)\;.
}
%\tgr{somewhere it would be good to have a commnet, perhps only a footnote, that a development using a $\mathbb{Z}$ grading is possible using shifts in graing rather than parity reversal}
We only distinguish even and odd elements, i.e. use a $\mathbb{Z}_2$-grading\footnote{The use of a $\mathbb{Z}$ is possible by using grading shifts, e.g. \cite{MR1958835}}. The observation \eqref{observe} is used to translate properties characterizing Lie (bi-)al\-ge\-broids from the differential graded to a more algebraic setting. In particular, if the anchor and bracket on $A$ are determined on a basis of sections $e_i$ of $A$  and $\partial_j$ of $TM$ by
\eq{
a(e_i) =\; a^j_i \partial_j\;, \quad [e_i,e_j]_A =\; f^k_{\,ij}e_k\;,
}
there is a  derivation $d_A$ on $\Gamma(\wedge^\bullet A^*)$ which translates into a vector field on $\Pi A$. Denoting the local coordinates on the latter by $(x^i,\xi^j)$, where $\xi^j$ denote the Grassmann generators, it is given by 
\eq{
d_A = \; a^j_i(x)\xi^i \partial_j - \frac{1}{2}f^k_{\,ij}(x)\xi^i\xi^j \frac{\partial}{\partial \xi^k}\;.
}
Analogously, if the dual $A^*$ is a Lie algebroid with anchor and bracket on a basis $e^i$ expressed by
\eq{
a_*(e^i) =\; a^{ij}\partial_j\;, \quad [e^i,e^j]_{A^*} =\; Q_k^{\,ij}(x)e^k\;
}
and using $(x^i,\theta_j)$ as local coordinates and generators on $\Pi A^*$, the differential $d_{A^*}$ is given by 
\eq{
d_{A^*} =\; a^{ij}(x)\theta_i\partial_j - \frac{1}{2} Q_k{}^{\,ij}(x)\theta_i\theta_j\frac{\partial}{\partial \theta_k}\;.
}
In the case of Lie algebroids $A$ and $A^*$, there are  derivations $d_A$ and $d_{A^*}$ which square to zero. In terms of the graded commutator of vector fields, this means $[d_A,d_A] = 0$ and analogously for $d_{A^*}$, i.e. they are \emph{homological}. With this terminology, the definition of a Lie algebroid (following \cite{Deethesis}) can be given in the most compact form:
\begin{definition}
A \emph{Lie algebroid} is a vector bundle $A \rightarrow M$ together with a homological vector field $d_A$ of degree 1 on the supermanifold $\Pi A$.
\end{definition}
The notation $f$ and $Q$ for the structure constants determined by the brackets on $A$ and $A^*$, respectively is common in the string theory literature \cite{Shelton:2005cf, Shelton:2006fd}, where these quantities are often called $f$- and $Q$-flux. We will use this nomenclature in the following. 
 
\subsubsection{Lie bialgebroids and the Drinfel'd double}
A pair $(A,A^*)$ of dual Lie algebroids is called \emph{Lie bialgebroid} if the differential $d_A$ is a derivation of the bracket on $A^*$ \cite{0885.58030} or, equivalently, if the differential $d_{A^*}$ is a derivation of the bracket on $A$.  In order to give an elegant characterization of this statement and for applications to DFT, it is desirable to have a meaningful sum of the two operators $d_A$ and $d_{A^*}$. A priori, they act on different spaces but in considering the cotangent bundles of $\Pi A$ and $\Pi A^*$, the function on $T^*\Pi A^*$ corresponding to $d_{A^*}$ can be pulled back by a Legendre transform to $T^*\Pi A$ and then added to the function corresponding to $d_A$. 

More precisely, let us extend the coordinates on $\Pi A$ by their canonically conjugate momenta on $T^*\Pi A$, denoted by a superscript star and a lower index, we get $(x^i,\xi^j,x^*_i,\xi^*_j)$ as coordinate set. In other words, the canonical Poisson relations on $T^*\Pi A$ are 
\eq{
\{x^i,x^*_j\} = \;\delta^i_j\;, \quad \{\xi^i,\xi^*_j\} =\;\delta^i_j \;,
}  
and vanishing Poisson brackets for the other combinations. Similar definitions are made for $T^*\Pi A^*$, whose coordinates should be denoted by $(x^i,\theta_j,x^*_i,\theta_*^j)$. It turns out \cite{Deethesis} that there is a symplectomorphism $L: T^*\Pi A \rightarrow T^*\Pi A^*$, called \emph{Legendre transform}, which relates the parity reversed fibre coordinates $\xi^i$ on $T^*\Pi A$ to the conjugate momenta $\theta_*^i$ on $T^*\Pi A^*$. More precisely we have:
\eq{
\label{Legendre}
L(x^i,\xi^j,x^*_i,\xi^*_j) =\; (x^i, \xi^*_j, x^*_i, \xi^j)\;.
}
Thus, if we denote the canonical projections by $\mathfrak{p}: T^*\Pi A \rightarrow \Pi A$ and $\bar{\mathfrak{p}}: T^*\Pi A^* \rightarrow \Pi A^*$, we have the following situation:
\eq{
\label{diagram}
\begin{matrix}
T^*\Pi A & \overset{L}\rightarrow & T^*\Pi A^* \\
\downarrow \mathfrak{p} & & \downarrow \bar{\mathfrak{p}} \\
\Pi A & & \Pi A^* 
\end{matrix}
}
Using the Legendre transform, it is possible to lift both differentials to $T^*\Pi A$. Expressed in the coordinates on the latter, the two functions are given by:
\eq{
\label{differentials}
  h_{d_A} =& \;  a^j_i(x) x_j ^*\xi^i -\frac{1}{2} f^k_{\,ij}(x)\xi^i\xi^j \xi^*_k \\
  L^* h_{d_{A^*}} =&\; a^{ij}(x)x_i^* \xi_j^* - \frac{1}{2} Q_k^{\,ij}(x)\xi^k\xi_i^*\xi_j^*.
}

The partial derivative operators in (3) and (5) are  realized by taking Poisson brackets with the functions $h_{d_A}$ and $L^* h_{d_{A^*}}$, respectively. It is now possible to add the two functions as they are defined on the same domain. The sum will turn out to be useful in order to characterize Lie bialgebroids and to reveal the structure of the C-bracket of Double Field Theory. Define
\eq{
\theta = \; h_{d_A} + L^*h_{d_{A^*}}\;.
}

%\trd{why $\theta$ rather than his $\mu$?} \tpr{Wanted to reserve $\mu$ for later use, but lets decide after further writing}

The characterization of Lie bialgebroids using $\theta$ is very concise and elegant and is given by the following theorem, whose detailed proof can be found in \cite{Deethesis}:
\begin{theorem}
The pair $(A,A^*)$ of a Lie algebroid $A$ and its linear dual is a Lie bialgebroid if and only if $\{\theta,\theta\} = 0$.
\end{theorem}
Having in mind the Courant algebroid associated to a Lie bialgebroid, the theorem gives a transparent characterization of this class of Courant algebroids. Motivated by these results, the space $T^*\Pi A$ together with its structure is the basis for defining the Drinfel'd double of a Lie bialgebroid \cite{mackenzie:drinfeld, kirill:Crelle}.
\begin{definition}
For a Lie bialgebroid $(A,A^*)$, the space $T^*\Pi A$ together with the homological vector field $D = \{\theta, \cdot\}$ is called the \emph{Drinfel'd double} of $(A,A^*)$. 
\end{definition}
In the following we will review briefly the relevance of the Drinfel'd double for structures arising in DFT. A detailed derivation of the results is given in \cite{Deser:2014mxa}. Starting with a glance at the gauge algebra of DFT, mainly to set up notation in the next subsection, we will continue with the main result to be used in later chapters: The representation of the C-bracket in terms of Poisson brackets on the Drinfel'd double. 

%\tbl{somewhere we should recall we showed the C-bracket was in fact Courant...}

\subsection{DFT and $\alpha'$-deformations}
Interpreting the winding degrees of freedom of closed string theory as a new set of canonical momenta in addition to the standard momenta conjugate to configuration space coordinates is the starting point of \emph{Double Field Theory}, an attempt to formulate a target space field theory framework manifestly invariant under the action of T-duality. The action of DFT exhibits a gauge symmetry governed by the \emph{C-bracket} which is a DFT version of the Courant bracket of generalized geometry (and reduces to various forms of the latter by restricting the set of coordinates to a manifold of the dimension of the original configuration space). Recently, this structure was extended to the first order in $\alpha'$, in the sense of the derivative expansion of the closed string sigma model on the sphere (i.e. classical string theory). In the following two subsections, we pick out the most important facts of the gauge algebra of DFT for later parts of this work. We then give the results of the first order $\alpha'$-correction. In later chapters we will see that it is possible to interpret $\alpha'$ as a deformation parameter of a star product. This is intriguing as it may hint at an interpretation of closed string sigma model perturbation theory as a star product expansion, which is already known for the case of topological sigma models \cite{Cattaneo:1999fm}.
 
\subsubsection{The gauge algebra of DFT}
Closed string sigma models on toroidal target spaces, i.e. $X : \Sigma \rightarrow M=T^d$, where $\Sigma$ is homeomorphic to $S^1 \times \mathbb{R}$ and $T^d$ is the $d$-dimensional torus, exhibit the special property of having two different sets of momenta. Including a $B$-field $B\in \Gamma(\wedge^2 T^*M)$, the action reads 
\eq{
\label{sigma}
S = \; \int_\Sigma \Bigl(h^{\alpha \beta}\partial_\alpha X^i \partial_\beta X^j G_{ij})\,d\mu_\Sigma + \int_\Sigma \,X^*B \;,
} 
where the two-dimensional metric is $h = \textrm{diag}(-1,1)$ and $X^*B$ denotes the pullback of $B$ to the worldsheet $\Sigma$. 

%\trd{I don't understand  the anti-symmetric $\epsilon$ - why ca't this be written in the usual pullback formalism?}

The target space metric is denoted by $G \in \Gamma(\otimes^2 T^*M)$. Solving the classical equations of motion to \eqref{sigma} for constant metric and $B$-field with appropriate boundary conditions leads to the well-known mode-expansions of the sigma model fields $X^i(\tau, \sigma)$, with $(\tau, \sigma)$ denoting the coordinates on $\Sigma$:
\eq{
\label{classic}
X_R^i =& x_{0R}^i + \alpha_0^i(\tau - \sigma) + i\underset{n\neq 0}\sum \frac{1}{n} \alpha_n^i e^{-in(\tau - \sigma)} \;,\\
X_L^i =& x_{0L}^i + \bar \alpha_0^i(\tau + \sigma) + i\underset{n\neq 0}\sum \frac{1}{n} \bar \alpha_n^i e^{-in(\tau +\sigma)}\;,
}
with integration constants $x_{0R},x_{0L}$, constant oscillator coefficients $\alpha_n^i, \bar{\alpha}_n^i, n\neq 0$ and zero modes $\alpha_{0}^i$ and $\bar{\alpha}_0^i$ given by:
\eq{
\alpha_0^i =& \frac{1}{\sqrt{2}}G^{ij}\Bigl(p_j - (G_{jk} + B_{jk})w^k\Bigr)\;,\\
\bar\alpha_0^i =&\frac{1}{\sqrt{2}} G^{ij}\Bigl(p_j + (G_{jk} - B_{jk})w^k\Bigr)\;.
}
The parameters $p_j$ correspond to the standard canonical momenta determined by the variational derivative of the action with respect to $\partial_\tau X^i$. In addition, we have the \emph{winding} parameters $w^i$ defined by $w^i = \frac{1}{2\pi}\int_{S^1} \,\partial_\sigma X^i \,d\sigma$, whose origin is the fact that we are considering closed string sigma models, i.e. for fixed $\tau$ we have a map $X: S^1 \rightarrow M$. 

The canonical momentum zero modes $p_i$ are usually interpreted as canonically conjugate to a set of coordinates $x^i$ on the target space\footnote{To distinguish coordinates on the manifold $M$ from the sigma model maps $X^i$, we denote them by small letters $x^i$.} $M$. 

%\trd{I'm bothered by M as a manifold and M as an index.}
%\tgr{how about P, Q instead of M, N??}

Following a similar interpretation for the winding zero modes $w^i$, one is lead to a second set of coordinates, usually denoted by $\tilde x_i$, in the sense that we have the correspondences:
\eq{
\label{operators}
p_i \simeq \frac{1}{i}\frac{\partial}{\partial x^i}\;, \quad \textrm{and} \quad w^i \simeq \frac{1}{i}\frac{\partial}{\partial \tilde x_i}\;.
}

%\trd{what does $simeq$ denote here?} \tla{It means ``corresponds to''. E.g. in quantum mechanics, the momentum coordinate corresponds to the derivative w.r.t. the position coordinate. In quantum mechanics they are related by Fourier transform.}

Taking this formal ``doubling'' of the configuration space as the basis to set up a classical field theory framework is the idea of DFT. We only mention this motivation and refer the reader to the huge amount of literature on this fast growing field mentioned in the introduction and references therein. In the following, we pick out results and terminology of DFT which are important for the rest of this work without giving the proofs. We hope the reader is still able to follow the small amount of background in DFT which is needed for later chapters. 

A crucial question is how to make contact with standard classical field theory, i.e. how to reduce the doubled configuration space to a physically relevant subspace. It turns out that the level-matching constraint of closed string theory, rewritten in terms of the derivative operators $\partial_i$ and $\tilde \partial^i,$ leads to the right reduction called \emph{strong constraint} of DFT \cite{Siegel:1993th, Hohm:2010jy, Betz:2014aia}. Taking functions depending on both sets of coordinates $\phi(x^i, \tilde x_i)$ and $\psi(x^i,\tilde x_i)$, the strong constraint reads
\eq{
\partial_i\phi (x^i,\tilde x_i)\tilde \partial^i \psi(x^i,\tilde x_i) + \tilde\partial^i \phi(x^i,\tilde x_i) \partial_i \psi(x^i, \tilde x_i) = 0\;,
}

It turns out that it is possible to define doubled vector fields $V$ determined by the pair $V^M := (V^i(x^k,\tilde x_k),V_i(x^k,\tilde x_k))$ together with a generalized Lie derivative ${\cal L}_V$ reducing correctly to the corresponding quantities in Hitchin's \emph{generalized geometry} by applying the strong constraint to the component functions. As an example, by setting $\tilde \partial^i = 0$ (which is an obvious solution to the strong constraint), $V$ with components $V^M$ correctly reduces to a section $V^k(x)\partial_k + V_k(x)dx^k$ of the generlized tangent bundle, locally isomorphic to $TM \oplus T^*M$. The notation $V^M$, where the capital index contains upper and lower indices, is to indicate that these objects transform in the fundamental representation of $O(d,d)$, where a matrix $A \in O(d,d)$ satisfies  the relation 
\eq{
A\eta A^t =\, \eta\;, \quad \eta:=\; \begin{pmatrix}
0 & id \\
id & 0
\end{pmatrix}\;,
}  
and $id$ is the $d$-dimensional identity matrix. Capital indices are raised and lowered by the bi-linear form $\eta$ and contractions are performed in the standard way by summing over both upper and lower indices. Derivatives with upper indices are defined to be with respect to $\tilde x_i$, e.g. $V^K \partial_K \phi = V^k\partial_k\phi + V_k \tilde \partial^k \phi$. We will call the bi-linear form $\eta$ a \emph{metric} in the following and denote it by $\langle \cdot,\cdot\rangle$. More precisely, for $V = (V^i,V_i)$ and $W=(W^i,W_i)$ we have 
\eq{
\label{pair}
\langle V, W \rangle = \; V^P W^Q \eta_{PQ} = \; V^i W_i + V_i W^i\;.
}
One of the main results of DFT is the formulation of an action for a generalized metric and a generalized dilaton, invariant under $O(d,d)$ and reducing to the standard bosonic Neveu-Schwarz action of closed string theory by solving the strong constraint. As a further result, the action of DFT is invariant under a gauge symmetry which is determined by the generalized Lie derivative. The action of the latter on doubled scalars $\phi$ and doubled vectors with components $W^M$ is given by
\eq{
{\cal L}_V \phi =& V^K \partial_K \phi \;,\\
({\cal L}_V W)_K =& V^P\partial_P W_K + (\partial_K V^P -\partial^P V_K)W_P \;,\\
({\cal L}_V W)^K =& V^P\partial_P W^K -(\partial_P V^K -\partial^K V_P)W^P\;.
}
Similar to standard Riemannian differential geometry, the commutator of two generalized Lie derivatives gives a generalized Lie derivative with respect to the bracket which determines the structure of the gauge algebra:
\eq{
\Bigl[{\cal L}_{V},{\cal L}_{W}\Bigr] = \; -{\cal L}_{[V,W]_C}\;,
}
where $[V,W]_C$ is the  \emph{C-bracket} of DFT and is given in components by the following expression:
\eq{
\label{CbracketKomp}
\Bigl([V,W]_C\Bigr)^P = V^K\partial_K W^P - W^K\partial_K V ^P -\frac{1}{2}\Bigl(V^K\partial^P W_{K} - W ^K\partial^P V_{K}\Bigr)\;,
}
Separating vector and form parts according to $V= v + \gamma$ and $W = w + \omega$, this means e.g. for the bracket of $v$ and $\omega$:

\eq{
\label{Cbracket}
%\Bigl([X,Y]_C\Bigr)_i =&\; 0\;, \\
%\Bigl([X,Y]_C \Bigr)^i =\; X^k\partial_k Y^i - Y^k\partial_k X^i \;, \\
\Bigl([v,\omega]_C\Bigr)_i =&\; \;\;v^k\partial_k\omega_i -\frac{1}{2}(v^k\partial_i\omega_k - \omega_k\partial_i v^k) \;,\\
\Bigl([v,\omega]_C\Bigr)^i =&\;-\omega_k \tilde\partial^k v^i - \frac{1}{2}(v^k\tilde\partial^i\omega_k - \omega_k\tilde\partial^i v^k) \;.
%\Bigl([\eta, Y]_C \Bigr)_i =&\; -Y^k\partial_k \eta_i + \frac{1}{2}\Bigl(Y^k\tilde\partial^i\eta_k - \eta_k\tilde \partial^i Y^k \Bigr)\;, \\
%\Bigl([\eta,Y]_C \Bigr)^i =&\; \;\;\eta_k\tilde \partial^k Y^i + \frac{1}{2}\Bigl(Y^k\tilde\partial^i\eta_k - \eta_k\tilde\partial^i Y^k\Bigr)\;, \\
%\Bigl([\eta,\omega]_C\Bigr)_i =& \; \;\;\eta_k\tilde \partial^k \omega_i - \omega_k\tilde \partial^k \eta_i\;, \\
%\Bigl([\eta,\omega]_C\Bigr)^i = 0\;.
}
It is this bracket structure in  which we are interested. After this glance at DFT, we will review an earlier result \cite{Deser:2014mxa} on how it is possible to rewrite the bracket \eqref{CbracketKomp} in terms of Poisson brackets on the Drinfel'd double of a Lie bialgebroid. To complete our glance into DFT, we will then review a \emph{deformation} of the C-bracket by considering the first order correction in the string coupling $\alpha'$. %Using the representation of the C-bracket in terms of Poisson brackets we are then able to formulate our main result: The identification of the first order correction in $\alpha'$ by an expansion of the Poisson brackets with respect to a Moyal-Weyl star product with respect to an appropriate Poisson structure.

\subsubsection{$\alpha'$-deformed metric and C-bracket} 
The parameter $\alpha'$ of string theory is related to the fundamental string length $l_s$ by $\alpha' = l_s^2$. Taking the closed string sigma model, perturbative expansions are power series in $\alpha'$. Taking the renormalization group equations corresponding to this perturbative expansion gives (to lowest order in $\alpha '$) the Einstein equations together with equations for the $H$-flux $H = dB$ and dilaton. The effective target space field theory corresponding to these equations is given by the supergravity action
\eq{
\label{stringaction}
S = \int \, d^dx \sqrt{-\textrm{det}G}e^{-2\phi}\Bigl[R + 4(\partial \phi)^2 - \frac{1}{12} H_{ijk}H^{ijk}\Bigr]\;,
}
where the dilaton is denoted by $\phi$. From the viewpoint of DFT, this action is reproduced by applying the solution $\tilde \partial^i = 0$ to the doubled target space action. T-duality in isometric directions is realized by the Buscher rules and the appropriate geometric framework to deal with T-duality in this situation is generalized geometry. 
Computing higher order corrections in $\alpha '$ to the renormalization group equations mentioned above leads to $\alpha '$ corrections to the action \eqref{stringaction}, and understanding the systematics of the $\alpha '$ expansion and its consequences on the action of T-duality and generalized geometry (e.g. corrections, or said differently, \emph{deformations},to the Courant bracket) are one of the outstanding questions of contemporary string theory. Recently, by analysing $\alpha'$-deformed Lorentz-transformations of the $B$-field motivated by Green-Schwarz anomaly cancellation in heterotic string theory \cite{Hohm:2014eba, Hohm:2014xsa}, an $\alpha'$-infinitesimal deformation of the C-bracket \eqref{CbracketKomp} was found (and by applying the strong constraint, of the Courant bracket). It is given for doubled vectors $X = (X^i(x,\tilde x),X_i(x,\tilde x))$ and $Y = (Y^i(x,\tilde x),Y_i(x,\tilde x))$ by:
\eq{
\label{deformedCbracket}
[X,Y]_{\alpha'} := \; [X,Y]_C + \alpha ' \,[[X,Y]]\;,
}
where $[X,Y]_C$ is the standard C-bracket \eqref{CbracketKomp} and the first order deformation is given in components by:
\eq{
[[X,Y]]^K=\;-\Bigl(\partial^K\partial_Q X^P\partial_P Y^Q - X \leftrightarrow Y\Bigr)\;,
}
which means e.g. for the form part:
\eq{
[[X,Y]]_i =\; -\frac{1}{2}\Bigl(&\partial_i \partial_m X^n \partial_n Y^m + \partial_i \partial_m X_n \tilde \partial^n Y^m + \partial_i \tilde \partial^m X^n \partial_n Y_m  \\
&  + \partial_i \tilde \partial^m X_n \tilde \partial^n Y_m - X \leftrightarrow Y\Bigr)\;,
}
and similarly for the vector part $[[X,Y]]^i$. Furthermore, an $\alpha '$-deformation to the bilinear pairing \eqref{pair} was proposed in \cite{Hohm:2014eba} in such a way that the deformed pairing remains a scalar under infinitesimal transformations determined by the deformed C-bracket \eqref{deformedCbracket}. This deformation is given for $X,Y$ as above by
\eq{
\label{deformedinner}
\langle X,Y \rangle_{\alpha '}  :=\; \langle X,Y \rangle - \alpha ' \,\langle\langle X,Y\rangle \rangle
:= X^P Y_P -\alpha ' \, \partial_P X^Q\partial_Q Y^P\;.
}
Expanded in components, this can be written as:
\eq{
\label{deformcomponents}
\langle X,&Y\rangle_{\alpha '} =\; X^i\eta_i + \omega_i Y^i \\
&- \alpha ' \Bigl(\partial_m X^n \partial_n Y^m + \partial_m X_n \tilde \partial^n Y^m +\tilde \partial^m X^n\partial_n Y_m + \tilde \partial^m X_n \tilde \partial^n Y_m\Bigr)\;.
}
Applying the strong constraint, these deformations lead to deformed Courant algebroids (as both the bilinear pairing and the Courant bracket receive corrections).

 The main question of our work is about a systematic understanding of the form of the deformations \eqref{deformedCbracket} and \eqref{deformedinner} in terms of a star product expansion on an appropriate graded manifold. The main ingredient to proceed in this direction is the earlier result of a representation of the C-bracket in terms of Poisson brackets on the Drinfel'd double of a Lie bialgebroid.

%\trd{Those last two sentenced should be emphasized - perhaps include them also in the intro?}

%\tla{Yes. I mention this on the beginning of page 4 in the introduction. Is this ok?} YES

\subsection{The C-bracket in terms of Poisson brackets}
For an $n$-dimensional configuration space $M$, the observables (fields) of DFT formally\footnote{We say formally, because the physical configuration space is still $n$-dimensional and to make physical statements, one has to choose an $n$-dimensional polarization by solving the strong constraint.} depend on $2n$ variables, often denoted by $(x^a,\tilde x_a)$. Said differently, there are $2n$ differential operators $(\partial_a,\tilde \partial^a)$ acting on the dynamical fields of the theory. It is the latter viewpoint which enables us to make contact with the geometry of the Drinfel'd double: The two derivative operators $d_A$ and $d_{A^*}$, lifted to $T^*\Pi A$ suggest a canonical choice of two sets of \emph{momenta}:
\eq{
\label{differentials}
h_{d_A} =& \xi^i\Bigl(a^j_i(x) x_j^* -\frac{1}{2}f^k_{\,ij}(x)\xi^j\xi^*_k\Bigr) =:\xi^a p_a\;,\\  
h_{d_{A^*}} =& \xi^*_i\Bigl(a^{ij}x_j^* + \frac{1}{2}Q_k^{\,ij}\xi^k \xi^*_j\Bigr) =: \xi^*_a \tilde p^a\;.
}
As a consequence, we are able to associate to $p_a$ and $\tilde p^a$ derivative operators for functions $f \in {\cal C}^\infty(M)$, lifted to $T^*\Pi A$ in the following way:
\eq{
\label{derivatives}
\partial_a f := \;\{p_a,f\}\;,\quad \textrm{and}\quad \tilde \partial^a f :=\;\{\tilde p^a,f\}\;,
}
where we take Poisson brackets on $T^*\Pi A$. Note, that we also could introduce variables $(x^a,\tilde x_a)$ canonically conjugate to $(p_a,\tilde p^a)$ (by suitably adjusting the Poisson structure on $T^*\Pi A$ if necessary), but this step is not needed for the representation of the C-bracket in terms of Poisson brackets in the main theorem of this section, as for the C-bracket, only the derivative operators $(\partial_a,\tilde \partial^a)$ play a role.

To state the theorem, we first demonstrate how the lifts of vector fields and one-forms on $M$ to the Drinfel'd double look. 
Having the diagram \eqref{diagram} in mind, we define a projection $p: T^*\Pi A \rightarrow \Pi(A\oplus A^*)$ by $p^* X = L^*\bar{\mathfrak{p}}^* X$ for $X\in \Gamma(A)$ and $p^*\omega = \mathfrak{p}^*$ for $\omega \in \Gamma(A^*)$. Writing this out in components, we get for the lift of vector fields and one-forms:
\eq{
p^*(X^i e_i) =\; X^i\xi^*_i\;, \quad \textrm{and} \quad p^*(\omega_i e^i) = \;\omega_i\xi^i\;.
}
The following theorem gives a representation of the C-bracket of double field theory in terms of Poisson brackets on the Drinfel'd double $T^*\Pi A$: 
\begin{theorem}
\label{oldresult}
For sections $X + \eta$ and $Y+\omega$ of the direct sum $A \oplus A^*$ with lifts to $T^*\Pi A$ given by $\Sigma^1 = p^*(X + \eta)$ and $\Sigma^2 = p^*(Y+\omega),$ let the operation $\circ$ be defined by 
\eq{
\Sigma^1 \circ \Sigma^2 = \; \Bigl\{\{\xi^a p_a + \xi^*_a \tilde p^a, \Sigma^1\},\Sigma^2\Bigr\}\;.
}
Then the C-bracket in double field theory of $\Sigma^1$ and $\Sigma^2$ can be represented by:
\eq{
[\Sigma^1,\Sigma^2]_C =\; \frac{1}{2}\Bigl(\Sigma^1 \circ \Sigma^2 - \Sigma^2 \circ \Sigma^1 \Bigr)\;.
}
\end{theorem}
For a detailed proof, we refer the reader to \cite{Deser:2014mxa}. This representation of the C-bracket as a two-fold Poisson bracket shows that it can be seen as a derived structure in the sense of \cite{MR2223157, zbMATH02217997, KosmannSchwarzbach:2003en}. The fact that it can be written in terms of Poisson brackets will be of great importance for later chapters.

\hskip5ex If we have a star product, the first order in the deformation parameter of the star-commutator gives the Poisson bracket and higher orders can be seen as deformations of the Poisson bracket, which give corrections to the C-bracket as a consequence of the previous theorem.

\section{Graded Moyal-Weyl deformation of the metric and C-bracket}
\label{deformation}
One of the most immediate questions on the deformation of the metric and C-bracket (and, by taking a solution of the strong constraint, to the Courant bracket)
%\trd{application is to C-bracket but how much applies to Courant bracket?}
encountered in the last section is about an ordering principle to understand their precise form. Having a systematic explanation of the deformation at hand enables a treatment of questions about the uniqueness of the deformation and self-evidently opens up the possibility to calculate the next orders in the $\alpha'$-expansion which to our knowledge are not known up to now. Comparing the latter with independent calculations from DFT beyond the known orders of $\alpha'$ would be intriguing. Given the results reviewed in earlier sections, especially theorem \ref{oldresult}, the idea to get a systematic explanation of the deformation to first order in $\alpha '$ is not far to seek. Let us recall the definition \cite{Bayen:1977ha, Bayen:1977hb} of a \emph{formal}  star product\footnote{ $\mathcal{C}^\infty(M)[[t]]$ denotes formal power series in $t$ with smooth functions on $M$ as coefficients. We use the letter $t$ for the deformation parameter, which we later relate to $\alpha'$. }:
\begin{definition}
\label{stardef}
Let $(M,\pi)$ be a Poisson manifold and $f,g \in \mathcal{C}^\infty(M)$. A \emph{formal star product } $\star$ is a $\mathcal{C}^\infty(M)$-bilinear map
\eq{
\star: \mathcal{C}^\infty(M)[[t]] &\times \mathcal{C}^\infty(M)[[t]] \rightarrow \mathcal{C}^\infty(M)[[t]] \\
f &\star g =\;\underset{k=0}{\overset{\infty}{\sum}} \, t^k m_k(f,g) \;,
}
with bidifferential operators $m_k$ such that $\star$ has the following properties:
\begin{itemize}
\item $\star$ is associative.
\item $m_0(f,g) = fg$.
\item $m_1(f,g) - m_1(g,f) = \{f,g\}$.
\item $1\star f = f = f\star 1$.
\end{itemize}
\end{definition}
The second and third properties in the definition say that the star-commutator $f\star g - g\star f$ reproduces the Poisson bracket on $M$ in the first order of the deformation parameter. On the other hand, the higher order terms in the star commutator give higher order corrections to the Poisson bracket. Using this fact together with theorem \ref{oldresult}, we can systematically deform the C-bracket if we know the underlying star product. As we will see later, similar arguments hold for the deformation of the metric. 

Our choice of the star product is restricted to reproduce to first order the $\alpha'$ deformation of \emph{both} the metric and the C-bracket encountered in the previous section. We will see that such a star product exists but leave questions of uniqueness to further mathematical studies.

\subsection{Star commutators for graded Moyal-Weyl products}
In definition \ref{stardef}, the second and the third properties of the $m_k$ ensure that to first order in the deformation parameter, the \emph{star commutator} of smooth functions $f$ and $g$ reproduces the Poisson bracket on $M$ in the first order of the deformation parameter:
\eq{
\{f,g\} = \;m_1(f,g) - m_1(g,f) =\;\underset{t\rightarrow 0}\lim\; \frac{1}{t}\Bigl(f\star g - g\star f\Bigr)\;.
}

%\trd{compare this wording with that 2 paragraphs before?}
%\tpr{Yes, it is the same. I wanted to repeat it to emphasize in what sense we get ``corrections''. But feel free to change it appropriately.}

In other words,  dropping the limit in this equation gives a natural way of getting ``quantum corrections'' to the classical Poisson structure on $M$. We use the following notation:
\eq{
\{f,g\}^\star :=&\; f\star g - g\star f \\
=&\;\underset{k=1}{\overset{\infty}{\sum}} \, t^k\Bigl(m_k(f,g) - m_k(g,f)\Bigr) \\
=&\;\, \underset{k=1}{\overset{\infty}{\sum}}\, t^k\{f,g\}_{(k)}\;,
}
i.e. we denote the $k$-th order contribution to the Poisson bracket by $\{f,g\}_{(k)}$. As the simplest example, let $(M,\pi)$ be a Poisson manifold with constant Poisson structure $\pi$ and $f,g$ be smooth functions on $M$. Then the Moyal-Weyl star product $\star_M$ is given by
\eq{
f\star_M g =\; fg + t\;\pi^{ij}\partial_i f \partial_j g + \frac{t^2}{2}\, \pi^{ij}\pi^{mn} \partial_i\partial_m f \partial_j\partial_n g + \mathcal{O}(t^3)\;.
}

The full star product is given by an exponential series. In a purely mathematical context, other full formulas are known, though we are unaware of applications to physics. We observe
 that, in this case, the quadratic (or more generally the even) powers in the deformation parameter vanish for the star-commutator of $f$ and $g$, i.e. one has
\eq{
f\star g - g\star f =\; t\,\{f,g\} + \mathcal{O}(t^3)\;,
}
as for even powers, the Poisson tensors contribute with an even power of $(-1)$ when exchanging $f$ and $g$ and therefore the terms vanish in the commutator. As a consequence there would be no first order corrections to the Poisson bracket. This situation changes by considering \emph{graded} Poisson manifolds. As was already mentioned in the beginning, our choice of the manifold will be $T^*\Pi A$, which is a symplectic supermanifold. To get corrections to the Poisson bracket, one has to take the various signs of the graded context into account, in particular one has to take an appropriate graded star-commutator.

Let $I= i_1\cdots i_k$ and $J= j_1\cdots j_k$ with $\partial_I= \partial_{x^{i_1}}\cdots \partial_{x^{i_k}}$. A general expression for the star-commutator in the \emph{purely even} Moyal-Weyl case is given by
\eq{
\{f,g\}^\star = \underset{k=1}{\overset{\infty}{\sum}}\,t^k\Bigl(\underset{IJ}\sum\, m_k^{IJ}(\partial_If\partial_J g 
 - \partial_I g \partial_J f)\Bigr)\;.
}

\noindent where in the case at hand, the constant Poisson structure is not differentiated and hence collected in the constant factors $m_k^{I J}.$
The generalization of this expression to the graded case is determined by the Koszul sign convention, i.e. whenever exchanging two objects or maps, one introduces a sign, e.g.  $(-1)^{|f||g|}$. 
%Further signs are chosen in such a way to reproduce the physics results and to get the graded Leibniz rule for the Poisson bracket, i.e. for the first order of the star commutator. 
We are led to the following sign convention: 
\eq{
\label{superstarcomm}
\{f,g\}^\star =\; \underset{k=1}{\overset{\infty}{\sum}}\,t^k\Bigl(\underset{IJ}{\sum}\, m_k^{IJ} \bigl(\partial_I f \partial_J g -(-1)^{|f||g| + |x^J|(|f|-1) + |x^I|(|g|-1)} \partial_I g \partial_J f \bigr)\Bigr)\;,
}

Here we use the notation $|x^I|$ for the sum of the degrees of the $x^{i_n}$, i.e. we have $|x^I| := |x^{i_1}|+\dots + |x^{i_k}|$.
The graded star commutator \eqref{superstarcomm} clearly reduces to the standard star commutator in the purely even case. As we will see later, it is this sign prescription which gives the correct reproduction of the $\alpha'$-corrections encountered in the physics literature. 

%\tbl{Aha!}

Furthermore, using this sign convention, it is easy to show that the purely even and purely odd parts of the Poisson structure determined by the first order of the star-commutator obey the graded Leibniz rule (graded derivation rule), i.e. we have
\eq{
\label{gradedLeibniz}
\{f,gh\} =\; \{f,g\}\,h + (-1)^{|f| |g|}\,g\,\{f,h\}\;,
}
for functions $f$ and $g$ of degrees $|f|$ and $|g|$, respectively. This is the standard Leibniz rule for the graded context.

\subsection{The choice of Poisson structure}
Let now $M$ be a symplectic manifold with Poisson structure $\pi$, given by the inverse of the symplectic form. We specify $A = TM$ and thus $A^*= T^*M$. It can be readily checked that the pair $(A,A^*)$ in this case is a Lie bialgebroid. Recall that the differentials are given by \eqref{differentials}. In order to avoid exhausting calculations, we choose the simplest case which still shows the essential features. All the expressions for the deformed metric and C-bracket we encountered in previous chapters appeared without fluxes $f$ and $Q$, so we look for a setup where these two vanish. In particular, to have vanishing $f,$ we choose the standard basis of $A$ such that the anchor is $a^i_j = \delta^i_j$. Furthermore, to have vanishing $Q$, we take the Poisson structure on $M$ to be constant. Clearly this setup is very special, but, as we will see, it will reproduce the deformation correctly. The general case is much more involved and goes beyond the scope of this work. We briefly comment on the inclusion of non-vanishing fluxes in our conclusions. 

In the setup described so far, the lifted differentials $h_{d_A}$ and $L^*h_{d_{A^*}}$ are particularly easy to deal with:
\eq{
\label{reducediff}
h_{d_A} =\;\xi^m x^*_m =\; \xi^m p_m\;, \quad &\textrm{i.e.} \quad p_m =\; x^*_m\;, \\
L^* h_{d_{A^*}} =\; \xi^*_m \pi^{mn} x^*_n =\; \xi^*_m \tilde p^m\;, \quad &\textrm{i.e.} \quad \tilde p^m =\; \pi^{mn} x^*_n\;.
}
As a consequence of this setting, we have the following result for the derivative operator $\tilde \partial^i$, which we defined in \eqref{derivatives}:
\eq{
\label{tildepartial}
\tilde \partial^i f =\; \{\tilde p^i,f\} = \;\pi^{ij}\{x^*_j,f\} =\; \pi^{ij}\partial_j f\;.
}
This is a particular solution for the strong constraint. In the following, we will prove the deformation up to first order in $\alpha '$ for the Courant algebroid corresponding to this solution. We  remark about the general situation at the end of this subsection.
%\trd{?appears already in DFT?}

To complete the framework, we have to choose the Poisson structure on $T^*\Pi A$. We take the following:
\eq{
\label{Poissonstructure1}
\pi_{T^*\Pi A} =\; \frac{\partial}{\partial x^*_i}\wedge\frac{\partial}{\partial x^i} + \frac{\partial}{\partial \xi^*_i} \wedge \frac{\partial}{\partial \xi^i} + \frac{\partial}{\partial x^i}\wedge \frac{\partial}{\partial \xi^*_i} - \pi^{ij}\,\frac{\partial}{\partial x^i}\wedge \frac{\partial}{\partial \xi^j}\;.
}
In the following sections, we will justify this choice by computing the deformations of the C-bracket (or, more precisely, the corresponding Courant bracket) and the metric using the graded star commutator. For the star product, we will choose the graded generalization of the Moyal-Weyl product corresponding to the Poisson structure \eqref{Poissonstructure1}.

To compare with expressions of double field theory later on, we also give the Poisson structure using the derivative operator $\tilde \partial^i$. Having in mind \eqref{tildepartial}, we can rewrite $\tilde \partial^i$ as the derivative with respect to a coordinate $\tilde x^i$ and thus:
\eq{
\label{Poissonstructure2}
\pi_{T^*\Pi A} =\; \frac{\partial}{\partial x^*_i} \wedge \frac{\partial}{\partial x^i} + \frac{\partial}{\partial \xi^*_i}\wedge \frac{\partial}{\partial \xi^i} + \frac{\partial}{\partial x^i} \wedge \frac{\partial}{\partial \xi^*_i} + \;\frac{\partial}{\partial \tilde x_i}\wedge \frac{\partial}{\partial \xi^i}\;.
}
We will use this Poisson structure in our computations in order to get the $\alpha'$-deformations encountered in double field theory and described in the previous sections.
\begin{rem} Having a concrete realization of the coordinates $\tilde x^i$ as given for example in \cite{Vaisman:2012ke}, one could also take the Poisson structure \eqref{Poissonstructure2} as a starting point. In the following calculations and especially in the appendix, we will see that we reproduce the result of double field theory up to terms which are of the form $\partial_i \tilde \partial^i \phi(x,\tilde x)$, where $\phi$ is one of the fields involved. These terms are zero as a consequence of the strong constraint\footnote{The condition $\partial_i \tilde \partial^i \phi = 0$ is called the \emph{weak constraint} in DFT.} if one considers double field theory. In the special situation \eqref{tildepartial},
 these terms vanish trivially due to the anti-symmetry of the Poisson structure, so in both cases we will be able to reproduce the results obtained in physics.
\end{rem}

\subsection{Deforming the metric}
We now have all the ingredients to start with deforming the bilinear form $\langle,\rangle$, which we also call the metric. Already for the easy and special setup chosen in the last subsection, the computations are lengthy due to the formula for the Moyal-Weyl star product at second order in the deformation parameter. 

%\tgr{first deformation versus up to second order - meaning ignoring second order but the sentence seems to say the comutation is lengthy AT second order??}

%\tla{We need the second order of the Moyal-Weyl formula. In the star commutator the zeroth order drops out and the first order is the Poisson bracket, so the second order is the first non-trivial correction to the Poisson bracket.}

This is due to the graded Poisson structure \eqref{Poissonstructure2}, which contains also contributions for the odd variables. For convenience in reading, we give an explicit expression for the star product in the appendix and only give the important steps for the results in the main text. 

%\tbl{YES!}

Let $V= \,V^m(x,\tilde x)\xi^*_m + V_m(x,\tilde x)\xi^m$ and $W=\,W^m(x,\tilde x)\xi^*_m + W_m(x,\tilde x)\xi^m$ be the lifts of generalized vectors to $T^*\Pi A$. The dependence on the tilded coordinates is to remind us  that we have two different derivative operators.  To use the star commutator to get deformations of the metric, we first note that the pairing $\langle V,W \rangle$ can be expressed as a Poisson bracket (i.e. the first order of the graded star commutator) on $T^*\Pi A$, using the Poisson structure \eqref{Poissonstructure2}:
\eq{
2\{V,W\}=\;&\frac{\partial V}{\partial \xi^*_i}\frac{\partial W}{\partial \xi^i} + \frac{\partial V}{\partial \xi^i}\frac{\partial W}{\partial \xi^*_i} - (-1)^{1}\Bigl(\frac{\partial W}{\partial \xi^*_i}\frac{\partial V}{\partial \xi^i} + \frac{\partial W}{\partial \xi^i}\frac{\partial V}{\partial \xi^*_i}\Bigr) \\
+&\frac{\partial V}{\partial x^i}\frac{\partial W}{\partial \xi^*_i} - \frac{\partial V}{\partial \xi^*_i}\frac{\partial W}{\partial x^i} - (-1)^{1}\Bigl(\frac{\partial W}{\partial x^i}\frac{\partial V}{\partial \xi^*_i} - \frac{\partial W}{\partial \xi^*_i}\frac{\partial V}{\partial x^i}\Bigr) \\
+&\frac{\partial V}{\partial \tilde x_i}\frac{\partial W}{\partial \xi^i} - \frac{\partial V}{\partial \xi^i}\frac{\partial W}{\partial \tilde x_i} - (-1)^{1}\Bigl(\frac{\partial W}{\partial \tilde x_i}\frac{\partial V}{\partial \xi^i} - \frac{\partial W}{\partial \xi^i}\frac{\partial V}{\partial \tilde x_i}\Bigr)\\
=&\; 2(V^iW_i + V_i W^i) =\; 2\langle V,W\rangle\;.
}

%\trd{cancellation of terms is in each row? how do you see that next to last =?}\tpr{Yes, thanks for asking. The first line gives the non-vanishing part. In the second line, for example the first term is cancelled by the last. Even without plugging in V and W: In the last term the derivative of W w.r.t. $\xi^*_i$ has degree $0$ and thus commutes with the second factor.}

As a consequence, we can compute  the second order correction to the pairing $\langle\cdot,\cdot \rangle$ by computing the correction to the Poisson bracket on $T^*\Pi A$ using the result \eqref{superstarcomm}. Whereas the first order in $t$ gives the pairing itself, for the second order we get\footnote{For the detailed calculation, we refer the reader to the appendix.}:
\eq{
\{V,W\}_{(2)} = \;&-\partial_i V^j \partial_j W^i - \partial_i V_j\tilde \partial^j W^i - \tilde \partial^i V^j\partial_j W_i - \tilde \partial^i V_j \tilde \partial^j W_i \\ 
=\;&-\partial_M V^N \partial_N W^M\;.
}
But this is exactly the deformation encountered in DFT. Thus we can identify $t$ with the deformation parameter $\alpha'$ in DFT. 

%\tbl{Wow!!}
%\tgr{ I'm still having language problems in re: orders. above compute  the second order correction but in Theroem up to second order??}

\begin{theorem}
\label{result1}
Let $V = V^i\xi^*_i + V_i \xi^i$ and $W = W^i\xi^*_i + W_i\xi^i$ be the lifts of two generalized vectors to $T^*\Pi A$. Then we have 
\eq{
\frac{1}{t}\{V,W\}^\star =\; \langle V,W\rangle - t\,\langle \langle V,W\rangle \rangle + \mathcal{O}(t^2)\;,
}
i.e. the graded star-commutator of $V$ and $W$ gives the deformed inner product of double field theory up to second order.
\end{theorem}
As already mentioned, the proof is straight forward by expanding the star product up to second order in the deformation parameter. We give the details in the appendix and move on to the deformation of the C-bracket in the next section.

\subsection{Deforming the C-bracket}
\label{Cdeformation}
According to theorem \ref{oldresult}, it is possible to express the C-bracket of double field theory in terms of a two-fold Poisson bracket on $T^*\Pi TM$. A short calculation shows that, in the lowest non-trivial order in the deformation parameter, the theorem is true for the Poisson structure \eqref{Poissonstructure2} and the lowest order contribution to the star-commutator \eqref{superstarcomm}, i.e. we have the following, if we take the operation $\circ$ from theorem \ref{oldresult} and $V=V^m(x,\tilde x)\xi^*_m + V_m(x,\tilde x)\xi^m$ and $W= W^m(x,\tilde x)\xi^*_m + W_m(x,\tilde x)\xi^m$: 
\eq{
V\circ W = \Bigl\{ \{\theta, V\}_{(1)},W\Bigr\}_{(1)}\;.
}  
As we are considering a two-fold Poisson bracket, the lowest order non-trivial contributions to the two-fold star-commutator are of order $t^2$ and $t^3$. Expanding the two-fold star-commutator at these orders, we have:
\eq{
\label{expansion}
\Bigl\{\{\theta,V\}^*,W\Bigr\}^* = \; t^2\,V\circ W + t^3\,\Bigl\{\{\theta,V\}_{(2)},W\Bigr\}_{(1)} + t^3\,\Bigl\{\{\theta,V\}_{(1)},W\Bigr\}_{(2)} + {\cal O}(t^4)\;.
}
As in the previous section, we now list the results for the various contributions to the Poisson brackets and refer the reader to the appendix for detailed calculations, which are straight forward, but tedious.

%\tgr{Recall $_{(i)}$} \tla{?}

First we get
\eq{
\label{intermediate}
\{\theta,V\}_{(1)} =\; &\xi^m\xi^n\partial_m V_n + \xi^*_k\xi^m\pi^{kn}\partial_n V_m + \xi^n \xi^*_m \partial_n V^m + \xi^*_k\xi^*_m \pi^{kn}\partial_n V^m \\ 
&+ V_n\pi^{nm}x^*_m + x^*_n V^n\;, \\ 
\{\theta,V\}_{(2)} =\; &-2\partial_n \tilde \partial^n V = \; -2 \pi^{nm}\partial_n\partial_m V = 0\;.
}
Note that we have used the assumption that the Poisson structure $\pi^{nm}$ is constant and anti-symmetric. In double field theory, a term of the form $\partial_n \tilde \partial^n V$ would vanish due to the strong constraint. Having the previous result, we can compute the Poisson brackets with $W$ needed for \eqref{expansion}. The only remaining term is 
\eq{
\Bigl\{\{\theta, V\}_{(1)}&,W\Bigr\}_{(2)} =\\
&\; 2\xi^m\Bigl(\partial_k W^n \partial_m \partial_n V^k + \partial_k W_n \partial_m \tilde \partial^n V^k + \tilde \partial^k W^n \partial_m\partial_n V_k + \partial_k W_n \partial_m \tilde \partial^n V^k \Bigr)\\
&+ 2\xi^*_m \Bigl(\partial_k W^n \tilde \partial^m \partial_i V^k + \partial_k W_n \tilde \partial^m \tilde \partial^n V^k 
+ \tilde \partial^k W^n \tilde \partial^m \partial_n V_k + \tilde \partial^k W_n \tilde \partial^m \tilde \partial^n V_k \Bigr)\;.
}
We see that this is exactly of the form \eqref{deformedCbracket} encountered in double field theory and described in section \ref{review}. We therefore can formulate the following result: 

\begin{theorem}
\label{result2}
Let $V=V^i\xi^*_i + + V_i \xi^i$ and $W=W^i\xi^*_i + W_i \xi^i$ be the lifts of two generalized vectors to $T^*\Pi A$. Then we have
\eq{
\frac{1}{2\,t^2}\Bigl(\bigl\{\{\theta,V\}^*,W\bigr\}^* -  \bigl\{\{\theta,W\}^*,V\bigr\}^*\Bigr) = [V,W]_C + t\,[[V,W]]_C + {\cal O}(t^2)\;,
}
i.e. the two-fold graded star commutator of $V,W$ with $\theta$ coincides with the C-bracket of DFT up to second order in the deformation parameter.  
\end{theorem}

%\trd{have we earlier specified how the C-bracket of double field theory looks?}
In the last theorem -- to have a concise and suggestive notation -- we use the same letters for the generalized vector fields and their respective lifts to the Drinfel'd double. This should cause no confusion. 
As for theorem \ref{result1}, by looking at \eqref{deformedCbracket} we see that we can identify the deformation parameter $t$ with the square of the string lenth, i.e. $\alpha '$. For the proof we refer the reader to the calculations done in the appendix.

%\section{Inclusion of fluxes}
%\input{flux_7a}

\section{Conclusion and outlook}
The setup chosen in the previous chapter to calculate the deformation of the C-bracket (as a Courant bracket) and the bi-linear pairing of generalized geometry/DFT is very specific. First, instead of calculating the deformation for a specific solution to the strong constraint,  the Poisson structure \eqref{Poissonstructure2} could be taken as a starting point to explicitly calculate the deformation to the C-bracket in \emph{full} double field theory. To stay on a safe mathematical ground, the meaning of ``double manifolds'' has to be made precise.  

Secondly, we used a constant Poisson structure $\pi$ on the configuration space manifold $M$. The language used in this work has the advantage to be naturally exendable to more general Poisson structures by the notion of a \emph{twist}, introduced in \cite{Roytenberg:2001am}. Let $(M,\pi)$ be a Poisson manifold with constant $\pi$, and $\beta = \frac{1}{2}\, \beta^{ij}\partial_i \wedge \partial_j$ an arbitrary element of $\Gamma(\wedge^2 T^*M)$. Using the language of chapter \ref{review}, especially the Legendre transform \eqref{Legendre}, the lift of $\beta$ to the Drinfel'd double is given by the quadratic function:
\eq{
L^*\bar{\mathfrak{p}}^* \beta = \; \frac{1}{2}\,\beta^{ij}\xi^*_i \xi^*_j\;.
}
Using this, the \emph{twist} of the functions $\mu := h_{d_A}$ and $\gamma := L^*h_{d_{A^*}}$ in \eqref{differentials} is defined by 
\eq{
\mu_\beta :=\; \mu\;, \quad \gamma_\beta :=\; \gamma + X_\beta \,\mu\;, 
}
where the action of $X_\beta$ is defined by $X_\beta \,\mu := \,\{L^*\bar{\mathfrak{p}}^* \beta, \mu\}$. All the deformation calculations of the previous chapter should be extendable to the twisted quantities. It would be interesting to see how this changes the form of the C-bracket and its deformation. 

The twist used in this form could be of interest in physics, in particular DFT. Identifying the derivative operators $\partial_a $ and $\tilde \partial^a$ as introduced in \eqref{derivatives}, we can calculate the component form of the twisted derivative of $L^*\bar{\mathfrak{p}}^* \beta$: 
\eq{
\{L^* \gamma_\beta, L^*\bar{\mathfrak{p}}^* \beta\} =& \{\xi^*_a \tilde p^a + X_\beta \mu, L^* \bfrp \beta\} \\
=& \{\xi_a^* \tilde p^a, L^* \bfrp \beta\} + \bigl\{\{L^*\bfrp\beta,\mu\}, L^*\bfrp \beta\bigr\} \\
=& \{\xi_a^* \tilde p^a, L^* \bfrp \beta\} + L^*\bigl\{\{\bfrp \beta, (L^{-1})^* \mu\},\bfrp \beta\bigr\} \\
=& \frac{1}{2} \tilde \partial^a \beta^{bc}\xi_a^*\xi_b^* \xi_c^* + L^*\bfrp [\beta,\beta]_{SN} \\
=& \bigl(\frac{1}{2} \tilde \partial^a \beta^{bc} + \beta^{ma}\partial_m \beta^{bc}\bigr)\xi_a^*\xi_b^* \xi_c^* \;.
}
where we used the fact that the Legendre transformation is a symplectomorphism and the derived form of the Schouten-Nijenhuis bracket $[,]_{SN}$, given e.g. in \cite{Deethesis}. It is intriguing that the result of this short calculation coincides with the lift of the component expression for the $R$ flux as it is used in DFT, e.g. in \cite{Andriot:2012an}. 

Besides extending the framework of our work by flux-type objects as sketched above, there is the question about higher orders of $\alpha '$. To our knowledge, an extension of the C-bracket and the bilinear form of DFT beyond the first non-trivial order of $\alpha'$ is not known up to now. So calculating the next order of our deformation could give hints how which an extension might look like. Moreover, a comparison of different approaches to star products containing flux type objects, such as \cite{Mylonas:2013jha}, would be an interesting task. Finally, we plan to compare our results to the more algebraic approaches to deformation theory of Courant algebroids in the mathematics literature, e.g. using the Rothstein algebra as in \cite{MR3320226}.

\vspace{40pt}

%\emph{Acknowledgements:}  The authors wish to thank Dimitry Roytenberg, Marco Zambon, Barton Zwiebach and Olaf Hohm for discussions. 
\newpage

\begin{appendices}
\section{Details of the star-product calculations}
In this appendix, in order to make the different sign conventions as transparent as possible for the reader, we give the detailed calculations leading to the results given in the main text of this article. These are especially theorem \ref{result1} giving the deformation of the inner product up to second order in $\alpha'$ and theorem \ref{result2} containing the deformation of the C-bracket up to second order in $\alpha'$. Let us recall the Poisson structure on $T^*\Pi A$ used in the main text, 
\eq{
\label{Poissonappendix}
\pi_{T^*\Pi A} =\; \frac{\partial}{\partial x^*_i} \wedge \frac{\partial}{\partial x^i} + \frac{\partial}{\partial \xi^*_i}\wedge \frac{\partial}{\partial \xi^i} + \frac{\partial}{\partial x^i} \wedge \frac{\partial}{\partial \xi^*_i} + \;\frac{\partial}{\partial \tilde x_i}\wedge \frac{\partial}{\partial \xi^i}\;.
}
To write down explicit expressions for the star product and consequently for the deformation of the metric and C-bracket, we use the Moyal-Weyl formula up to second order in the deformation parameter, given in terms of the components $\pi^{ij}$ of the (constant) Poisson structure by
\eq{
f\star_M g =\; fg + t\;\pi^{ij}\partial_i f \partial_j g + \frac{t^2}{2}\, \pi^{ij}\pi^{mn} \partial_i\partial_m f \partial_j\partial_n g + \mathcal{O}(t^3)\;.
}
Applied to \eqref{Poissonappendix}, we get the expansion for the star product which we use for computations done in the following and referred to in the main text. Up to second order in the deformation parameter, we have
{\footnotesize
\eq{
\label{starproduct}
&f \star g = fg + t\Bigl[\bigl(\frac{\partial f}{\partial x^*_i}\frac{\partial g}{\partial x^i} - \frac{\partial f}{\partial x^i}\frac{\partial g}{\partial x^*_i}\bigr) + \bigl(\frac{\partial f}{\partial \xi^*_i}\frac{\partial g}{\partial \xi^i} + \frac{\partial f}{\partial \xi^i}\frac{\partial g}{\partial \xi^*_i}\bigr)  \\ 
&+ \bigl(\frac{\partial f}{\partial x^i}\frac{\partial g}{\partial \xi^*_i} - \frac{\partial f}{\partial \xi^*_i}\frac{\partial g}{\partial x^i}\bigr) + \bigl(\frac{\partial f}{\partial \tilde x_i}\frac{\partial g}{\partial \xi^i} - \frac{\partial f}{\partial \xi^i}\frac{\partial g}{\partial \tilde x_i} \bigr)\Bigr] \\
&+ \tfrac{t^2}{2}\Bigl[\bigl(\pxsi\pxsj f \pxi\pxj g - \pxsi \pxj f \pxi \pxsj g - \pxi \pxsj f \pxsi \pxj g + \pxi \pxj f \pxsi \pxsj g\bigr) \\
&+\bigl(\pxisi \pxisj f \pxii\pxij g  + \pxisi \pxij f \pxii \pxisj g + \pxii\pxisj f \pxisi \pxij g + \pxii \pxij f \pxisi \pxisj g \bigr) \\
&+ \bigl(\pxi \pxj f \pxisi \pxisj g - \pxi \pxisj f \pxisi \pxj g - \pxisi \pxj f \pxi \pxisj g  + \pxisi \pxisj f \pxi \pxj g \bigr) \\
&+ \bigl(\ptxi \ptxj f \pxii \pxij g - \ptxi \pxij f \pxii \ptxj g - \pxii \ptxj f \ptxi \pxij g  + \pxii \pxij f \ptxi \ptxj g \bigr) \\
&+2\bigl(\pxsi \pxisj f \pxi \pxij g + \pxsi \pxij f \pxi \pxisj g - \pxi \pxisj f \pxsi \pxij g - \pxi \pxij f \pxsi \pxisj g \bigr) \\
&+2 \bigl(\pxsi \pxj f \pxi \pxisj g - \pxsi \pxisj f \pxi \pxj g - \pxi \pxj f \pxsi \pxisj g + \pxi \pxisj f \pxsi \pxj g \bigr)\\ 
&+2\bigl(\pxsi \ptxj f \pxi \pxij g - \pxsi \pxij f \pxi \ptxj g  - \pxi \ptxj f \pxsi \pxij g + \pxi \pxij f \pxsi \ptxj g \bigr) \\
&+\bigl(\pxi\ptxj f \pxisi \pxij g - \pxi \pxij f \pxisi \ptxj g - \pxisi \ptxj f \pxi \pxij g  + \pxisi \pxij f \pxi \ptxj g \bigr) \\
&+\bigr(\ptxi\pxj f \pxii \pxisj g - \ptxi \pxisj f \pxii \pxj g - \pxii \pxj f \ptxi \pxisj g  + \pxii \pxisj f \ptxi \pxj g \bigr) 
\Bigr]
}
}
By using the star-commutator \eqref{superstarcomm} in the main text, we will give the details of the calculations needed for the main results.

\subsection{The deformation of the metric}
As in the main text, let $V=V^i(x,\tilde x)\xi^*_i + V_i(x,\tilde x)\xi^i$ and $W= W^i(x,\tilde x)\xi^*_i + W_i(x,\tilde x)\xi^i$ be the lifts of generalized sections to $T^*\Pi A$. Then the star product \eqref{starproduct} gives the following result for the star-commutator \eqref{superstarcomm}, up to second order in the deformation parameter:
\eq{
\{V,W\}^\star =\;& \frac{t}{2}\Bigl[\frac{\partial V}{\partial \xi^*_i}\frac{\partial W}{\partial \xi^i} + \frac{\partial V}{\partial \xi^i}\frac{\partial W}{\partial \xi^*_i} + \frac{\partial V}{\partial x^i}\frac{\partial W}{\partial \xi^*_i} - \frac{\partial V}{\partial \xi^*_i}\frac{\partial W}{\partial x^i} + \frac{\partial V}{\partial \tilde x_i}\frac{\partial W}{\partial \xi^i} - \frac{\partial V}{\partial \xi^i}\frac{\partial W}{\partial \tilde x_i} \\
& +\frac{\partial W}{\partial \xi^*_i}\frac{\partial V}{\partial \xi^i} + \frac{\partial W}{\partial \xi^i}\frac{\partial V}{\partial \xi^*_i}  + \frac{\partial W}{\partial x^i}\frac{\partial V}{\partial \xi^*_i} - \frac{\partial W}{\partial \xi^*_i}\frac{\partial V}{\partial x^i} + \frac{\partial W}{\partial \tilde x_i}\frac{\partial V}{\partial \xi^i} - \frac{\partial W}{\partial \xi^i}\frac{\partial V}{\partial \tilde x_i}\Bigr] \\
+&\frac{t^2}{4}\Bigl[-\pxi \pxisj V \pxisi \pxj W -\pxisi \pxj V \pxi \pxisj W - \ptxi \pxij V \pxii \ptxj W \\
&-\pxii \ptxj V \ptxi \pxij W -\pxi \pxij V \pxisi \ptxj W - \pxisi \ptxj V \pxi \pxij W  \\
&- \ptxi \pxisj V \pxii \pxj W - \pxii \pxj V \ptxi \pxisj W - \pxi \pxisj W \pxisi \pxj V \\ 
&- \pxisi \pxj W \pxi \pxisj V - \ptxi \pxij W \pxii \ptxj V - \pxii \ptxj W \ptxi \pxij V \\
&-\pxi \pxij W \pxisi \ptxj V - \pxisi \ptxj W \pxi \pxij V - \ptxi \pxisj W \pxii \pxj V \\
&- \pxii \pxj W \ptxi \pxisj V \Bigr]\;.
}
We remark that the zeroth order in the deformation parameter has cancelled as is expected from the purely even cases. Furthermore we see that in the last expression, lots of terms cancel or add up, so that we finally get the following result, where we already used the form of $V$ and $W$:
\eq{
\{V,W\}^\star =\;& t\bigl(V^iW_i + V_i W^i\bigr) \\
-&\frac{t^2}{2}\bigl(\partial_i V^j\partial_j W^i + \partial_jV^i\partial_i W^j + \tilde \partial^i V_j \tilde \partial^j W_i + \tilde \partial^j V_i \tilde \partial^i W_j \\
& +\partial_i V_j \tilde \partial^j W^i + \tilde \partial^j V^i \partial_i W_j + \tilde \partial^i V^j \partial_j W_i + \partial_j V_i \tilde \partial^i W^j\bigr) \\
=\;& t \langle V, W \rangle + t^2 \langle \langle V, W \rangle\rangle \;, 
}
which is the result of theorem \ref{result1}.  

%\eq{
%V \star W = & V^mW^k\xi^*_m \xi^*_k + V_m W^k\xi^m \xi^*_k + V^m W_k \xi^*_m \xi^k + V_m W_k \xi^m \xi^k \\
%&+t\Bigl[ V^iW_i + V_i W^i + W^i\partial_i V^m \xi^*_m + W^i \partial_i V_m \xi^m - V^i\partial_i W^k \xi^*_k - V^i\partial_i W_k \xi^k \\
%& + W_i \tilde \partial^i V^m \xi^*_m + W_i \tilde \partial^i V_m \xi^m - V_m \tilde \partial^m W^k \xi^*_k - V_m \tilde \partial^m W_k \xi^k \Bigr] \\
%& + \frac{t^2}{2} \Bigl[ - \partial_i V^j \partial_j W^i - \partial_j V^i\partial_i W^j -\tilde \partial^i V_j \tilde \partial^j W_i - \tilde \partial^j V_i \tilde \partial^i W_j \\
%& - \partial_i V_j \tilde \partial^j W^i - \tilde \partial^j V^i \partial_i W_j - \tilde \partial^i V^j \partial_j W_i - \partial_j V_i \tilde \partial^i W^j \Bigr]\;.
%}

\subsection{Deformation of the C-bracket}
To show the deformation of the C-bracket given in the main text, let us now use \eqref{reducediff} to express the homological vector field $\theta$ in terms of the momenta. We have $\theta = \xi^i x^*_i + \xi^*_i \pi^{ij}x^*_j$, and we let $V$ and $W$ be the lifts of generalized vector fields as in the previous section. The first two non-vanishing orders of the star commutator $\{\theta, V\}^*$ are given by 
\eq{
t\,&\{\theta, V\}_{(1)} + t^2\,\{\theta,V\}_{(2)} \\
=&\,\frac{t}{2}\Bigl[\frac{\partial \theta}{\partial x^*_i}\frac{\partial V}{\partial x_i} + \frac{\partial \theta}{\partial \xi^*_i}\frac{\partial V}{\partial \xi^i} + \frac{\partial \theta}{\partial \xi^i} \frac{\partial V}{\partial \xi^*_i} - \frac{\partial \theta}{\partial \xi^*_i }\frac{\partial V}{\partial x^i} - \frac{\partial \theta}{\partial \xi^i}\frac{\partial V}{\partial \tilde x_i}  + (-1)^1\frac{\partial V}{\partial x^i}\frac{\partial \theta}{\partial x^*_i} \\
& -(-1)^1\frac{\partial V}{\partial \xi^*_i}\frac{\partial \theta}{\partial \xi^i} - (-1)^1\frac{\partial V}{\partial \xi^i}\frac{\partial \theta }{\partial \xi^*_i} - (-1)^{1}\frac{\partial V}{\partial x^i}\frac{\partial \theta}{\partial \xi^*_i} -(-1)^1\frac{\partial V}{\partial \tilde x_i}\frac{\partial \theta}{\partial \xi^i}\Bigr] \\
&+\frac{t^2}{4}\Bigl[2\pxsi\pxisj\theta \pxi\pxij V + 2\pxsi\pxij \theta \pxi \pxisj V - 2\pxsi \pxisj \theta \pxi \pxj V \\
& - 2 \pxsi \pxij \theta \pxi \ptxj V +2(-1)^1 \pxi \pxisj V \pxsi \pxij \theta  + 2(-1)^1\pxi\pxij V \pxsi \pxisj \theta \\
& +2(-1)^1 \pxi\pxj V \pxsi \pxisj \theta + 2(-1)^1\pxi \ptxj V \pxsi \pxij \theta \Bigr] \\
=&\, t\Bigl(\frac{\partial \theta }{\partial x^*_i}\frac{\partial V}{\partial x^i} + \frac{\partial \theta}{\partial \xi^*_i}\frac{\partial V }{\partial \xi^i} + \frac{\partial \theta }{\partial \xi^i}\frac{\partial V}{\partial \xi^*_i} \Bigr) -t^2\Bigl(\pxsi \pxisj \theta \pxi \pxj V + \pxsi\pxij \theta \pxi \ptxj V\Bigr) \\
=&\,t\Bigl(\xi^i\xi^m \partial_i V_m + \xi^*_k\xi^m \pi^{ki}\partial_i V_m + \xi^i\xi^*_m \partial_i V^m + \xi^*_k \xi^*_m \pi^{ki}\partial_i V^m + V_i \pi^{ik} x^*_k + x^*_i V^i\Bigr) \\
&-t^2\Bigl(\partial_i\tilde \partial^i (V^m\xi^*_m + V_m \xi^m)\Bigr)\;.
}   
These are the results \eqref{intermediate} listed in the main part of the text. As was explained there, the only non-vanishing part contributing to the deformation of the C-bracket at order $t$ is given by $\{\{\theta, V\}_{(1)},W\}_{(2)}$, which we now compute in detail. To shorten notation, let us use $U:= \{\theta,V\}_{(1)}$. We remark, that $U$ depends on all variables involved and therefore the second order of the star commutator contains much more terms. After expanding the latter and collecting the terms which do not cancel pairwise from the two parts of the star commutator, we get
\eq{
\{U,W\}_{(2)} =&\,-\frac{1}{2}\Bigl(\pxi \pxisj U \pxisi \pxj W + \pxisi \pxj U \pxi \pxisj W + \ptxi \pxij U \pxii \ptxj W \\
& + \pxii \ptxj U \ptxi \pxij W +  \pxisi\ptxj U \pxi \pxij W +  \pxi \pxij U \pxisi \ptxj W \\
& + \ptxi \pxisj U \pxii \pxj W + \pxii \pxj U \ptxi \pxisj W + \pxisi \pxij U \pxii\ptxj W  \\
& +\pxii \pxij U \pxisi \ptxj W 
 + \pxii \pxisj U \ptxi \pxij W + \pxii \pxij U \ptxi \pxisj W \Bigr)
}
Now we use the specific form of $U$ and $W$ and collect the vector field and form parts of the result, i.e. the terms which contract with $\xi^i$ and $\xi^*_i$, respectively. We get:
\eq{
\{U,W\}_{(2)} =&\,-\xi^m\Bigl(\partial_j W^i\partial_i (\pi^{jk}\partial_k V_m) - \partial_j W^i\partial_i \partial_m V^j + \tilde \partial^jW_i \tilde \partial^i \partial_j V_m - \tilde \partial^i \partial_m V_j  \\
&+ \partial_i W_j \tilde \partial^j (\pi^{ir}\partial_r V_m) - \partial_i W_j \tilde \partial^j \partial_m V^i + \tilde \partial^j W^i \partial_i \partial_j V_m - \tilde \partial^j W^i \partial_i \partial_m V_j\Bigr) \\
&-\xi^*_m\Bigl(\partial_j W^i \partial_i (\pi^{jk}\partial_k V^m)  -\partial_j W^i \partial_i (\pi^{mr}\partial_r V^j) - \tilde \partial^j W_i \tilde \partial^i (\pi^{mr}\partial_r V_j) \\ 
& + \tilde \partial^j W_i \tilde \partial^i \partial_j V^m + \partial_i W_j \tilde \partial^j (\pi^{ir} \partial_r V^m) - \partial_i W_j \tilde \partial^j (\pi^{mr} \partial_r V^i)  \\
&- \tilde \partial^j W^i \partial_i (\pi^{kr} \partial_r V_j) + \tilde \partial^j W^i \partial_i \partial_j V^k\Bigr) \\
=&\, \xi^m\Bigl(\partial_k W^n\partial_m\partial_n V^k + \partial_k W_n \partial_m \tilde \partial^n V^k + \tilde \partial^k W^n \partial_m \partial_n V_k + \partial_k W_n \partial_m \tilde \partial^n V^k\Bigr) \\
&\xi^*_m \Bigl(\partial_k W^n \tilde \partial^m \partial_i V^k + \partial_k W_n \tilde \partial^m \tilde \partial^n V^k + \tilde \partial^k W^n \tilde \partial^m \partial_n V_k + \tilde \partial^k W_n \tilde \partial^m \tilde \partial^n V_n\Bigr)\;.
}
In the last step, we used that the Poisson structure $\pi^{ij}$ is constant and anti-symmetric and the form \eqref{tildepartial} of the derivative operator $\tilde \partial^i$. The case for general Poisson structures is more involved and in particular contains terms proportional to the $Q$-flux mentioned in the main text. This general case goes beyond the scope of the present work, where our goal was to derive the deformation of the C-bracket without any contribution of the fluxes $f$ and $Q$. 

The last line of the previous calculations is the result mentioned in section \ref{Cdeformation}. Therefore we derived the ingredients necessary for the proof of theorem \ref{result2}, which finishes the appendix on the calculational details.

\end{appendices}

\bibliographystyle{hep}

\bibliography{mybig}

\newcommand{\etalchar}[1]{$^{#1}$}
\def\cprime{$'$}
\begin{thebibliography}{GMPW09}

\bibitem[AHL{\etalchar{+}}12]{Andriot:2012an}
D.~Andriot, O.~Hohm, M.~Larfors, D.~L{\"u}st and P.~Patalong, \textsl{
  {Non-Geometric Fluxes in Supergravity and Double Field Theory}},
\newblock Fortsch.Phys. \textbf{ 60}, 1150--1186 (2012), {1204.1979}.

\bibitem[BBLR14]{Betz:2014aia}
A.~Betz, R.~Blumenhagen, D.~L{\"u}st and F.~Rennecke, \textsl{ {A Note on the
  CFT Origin of the Strong Constraint of DFT}},
\newblock JHEP \textbf{ 1405}, 044 (2014), {1402.1686}.

\bibitem[BDL{\etalchar{+}}11]{Blumenhagen:2011ph}
R.~Blumenhagen, A.~Deser, D.~L{\"u}st, E.~Plauschinn and F.~Rennecke, \textsl{
  {Non-geometric Fluxes, Asymmetric Strings and Nonassociative Geometry}},
\newblock J.Phys. \textbf{ A44}, 385401 (2011), {1106.0316}.

\bibitem[BFF{\etalchar{+}}78a]{Bayen:1977ha}
F.~Bayen, M.~Flato, C.~Fronsdal, A.~Lichnerowicz and D.~Sternheimer, \textsl{
  {Deformation Theory and Quantization. 1. Deformations of Symplectic
  Structures}},
\newblock Annals Phys. \textbf{ 111}, 61 (1978).

\bibitem[BFF{\etalchar{+}}78b]{Bayen:1977hb}
F.~Bayen, M.~Flato, C.~Fronsdal, A.~Lichnerowicz and D.~Sternheimer, \textsl{
  {Deformation Theory and Quantization. 2. Physical Applications}},
\newblock Annals Phys. \textbf{ 111}, 111 (1978).

\bibitem[BFH{\etalchar{+}}14]{Blumenhagen:2013zpa}
R.~Blumenhagen, M.~Fuchs, F.~Ha{\ss}ler, D.~L{\"u}st and R.~Sun, \textsl{
  {Non-associative Deformations of Geometry in Double Field Theory}},
\newblock JHEP \textbf{ 1404}, 141 (2014), {1312.0719}.

\bibitem[BL14]{Bakas:2013jwa}
I.~Bakas and D.~L{\"u}st, \textsl{ {3-Cocycles, Non-Associative Star-Products
  and the Magnetic Paradigm of $R$-Flux String Vacua}},
\newblock JHEP \textbf{ 1401}, 171 (2014), {1309.3172}.

\bibitem[BP11]{Blumenhagen:2010hj}
R.~Blumenhagen and E.~Plauschinn, \textsl{ {Nonassociative Gravity in String
  Theory?}},
\newblock J.Phys. \textbf{ A44}, 015401 (2011), {1010.1263}.

\bibitem[Bus87]{Buscher:1987sk}
T.~Buscher, \textsl{ {A Symmetry of the String Background Field Equations}},
\newblock Phys.Lett. \textbf{ B194}, 59 (1987).

\bibitem[CF00]{Cattaneo:1999fm}
A.~S. Cattaneo and G.~Felder, \textsl{ {A Path integral approach to the
  Kontsevich quantization formula}},
\newblock Commun.Math.Phys. \textbf{ 212}, 591--611 (2000), {math/9902090}.

\bibitem[{Cou}90]{zbMATH00004959}
T.~J. {Courant}, \textsl{ {Dirac manifolds.}},
\newblock {Trans. Am. Math. Soc.} \textbf{ 319}(2), 631--661 (1990).

\bibitem[DS14]{Deser:2014mxa}
A.~Deser and J.~Stasheff, \textsl{ {Even symplectic supermanifolds and double
  field theory}},
\newblock (2014), {1406.3601}.

\bibitem[GMPW09]{Grana:2008yw}
M.~Grana, R.~Minasian, M.~Petrini and D.~Waldram, \textsl{ {T-duality,
  Generalized Geometry and Non-Geometric Backgrounds}},
\newblock JHEP \textbf{ 0904}, 075 (2009), {0807.4527}.

\bibitem[GMX14]{2014arXiv1410.3346G}
M.~{Gr{\"u}tzmann}, J.-P. {Michel} and P.~{Xu}, \textsl{ {Weyl quantization of
  degree 2 symplectic graded manifolds}},
\newblock {ArXiv e-prints} \textbf{ {math.DG}} ({2014}), {{1410.3346}}.

\bibitem[GPR94]{Giveon:1994fu}
A.~Giveon, M.~Porrati and E.~Rabinovici, \textsl{ {Target space duality in
  string theory}},
\newblock Phys.Rept. \textbf{ 244}, 77--202 (1994), {hep-th/9401139}.

\bibitem[Gua03]{Gualtieri:2003dx}
M.~Gualtieri, \textsl{ {Generalized complex geometry}},
\newblock (2003), {math/0401221}.

\bibitem[HHZ10a]{Hohm:2010jy}
O.~Hohm, C.~Hull and B.~Zwiebach, \textsl{ {Background independent action for
  double field theory}},
\newblock JHEP \textbf{ 1007}, 016 (2010), {1003.5027}.

\bibitem[HHZ10b]{Hohm:2010pp}
O.~Hohm, C.~Hull and B.~Zwiebach, \textsl{ {Generalized metric formulation of
  double field theory}},
\newblock JHEP \textbf{ 1008}, 008 (2010), {1006.4823}.

\bibitem[Hit03]{Hitchin:2004ut}
N.~Hitchin, \textsl{ {Generalized Calabi-Yau manifolds}},
\newblock Quart.J.Math.Oxford Ser. \textbf{ 54}, 281--308 (2003),
  {math/0209099}.

\bibitem[HT87]{Hull:1987pc}
C.~Hull and P.~Townsend, \textsl{ {The Two Loop Beta Function for $\sigma$
  Models With Torsion}},
\newblock Phys.Lett. \textbf{ B191}, 115 (1987).

\bibitem[Hul07]{Hull:2007zu}
C.~Hull, \textsl{ {Generalised Geometry for M-Theory}},
\newblock JHEP \textbf{ 0707}, 079 (2007), {hep-th/0701203}.

\bibitem[HZ09a]{Hull:2009mi}
C.~Hull and B.~Zwiebach, \textsl{ {Double Field Theory}},
\newblock JHEP \textbf{ 0909}, 099 (2009), {0904.4664}.

\bibitem[HZ09b]{Hull:2009zb}
C.~Hull and B.~Zwiebach, \textsl{ {The Gauge algebra of double field theory and
  Courant brackets}},
\newblock JHEP \textbf{ 0909}, 090 (2009), {0908.1792}.

\bibitem[HZ14a]{Hohm:2014xsa}
O.~Hohm and B.~Zwiebach, \textsl{ {Double Field Theory at Order $\alpha'$}},
\newblock (2014), {1407.3803}.

\bibitem[HZ14b]{Hohm:2014eba}
O.~Hohm and B.~Zwiebach, \textsl{ {Green-Schwarz mechanism and
  $\alpha'$-deformed Courant brackets}},
\newblock (2014), {1407.0708}.

\bibitem[Ket00]{Ketov:2000dy}
S.~Ketov,
\newblock \textsl{ {Quantum nonlinear sigma models: From quantum field theory
  to supersymmetry, conformal field theory, black holes and strings}},
\newblock {Springer, Berlin}, 2000.

\bibitem[KM97]{Kaloper:1997ux}
N.~Kaloper and K.~A. Meissner, \textsl{ {Duality beyond the first loop}},
\newblock Phys.Rev. \textbf{ D56}, 7940--7953 (1997), {hep-th/9705193}.

\bibitem[KS04]{KosmannSchwarzbach:2003en}
Y.~Kosmann-Schwarzbach, \textsl{ {Derived brackets}},
\newblock Lett.Math.Phys. \textbf{ 69}, 61--87 (2004), {math/0312524}.

\bibitem[KW15]{MR3320226}
F.~Keller and S.~Waldmann, \textsl{ Deformation theory of {C}ourant algebroids
  via the {R}othstein algebra},
\newblock J. Pure Appl. Algebra \textbf{ 219}(8), 3391--3426 (2015).

\bibitem[L{\"u}s10]{Lust:2010iy}
D.~L{\"u}st, \textsl{ {T-duality and closed string non-commutative (doubled)
  geometry}},
\newblock JHEP \textbf{ 1012}, 084 (2010), {1010.1361}.

\bibitem[LWX97]{0885.58030}
Z.~Liu, A.~Weinstein and P.~Xu, \textsl{ {Manin triples for Lie
  bialgebroids.}},
\newblock J. Differ. Geom. \textbf{ 45}(3), 547--574 (1997).

\bibitem[Mac98]{mackenzie:drinfeld}
K.~C.~H. Mackenzie, \textsl{ Drinfel\cprime d doubles and {E}hresmann doubles
  for {L}ie algebroids and {L}ie bialgebroids},
\newblock Electron. Res. Announc. Amer. Math. Soc. \textbf{ 4}, 74--87
  (electronic) (1998).

\bibitem[Mac05]{Mackenzie:GT}
K.~C.~H. Mackenzie,
\newblock \textsl{ General theory of {L}ie groupoids and {L}ie algebroids},
  volume 213 of \textsl{ London Mathematical Society Lecture Note Series},
\newblock Cambridge University Press, Cambridge, 2005.

\bibitem[Mac11]{kirill:Crelle}
K.~C.~H. Mackenzie, \textsl{ Ehresmann doubles and {D}rinfel'd doubles for
  {L}ie algebroids and {L}ie bialgebroids},
\newblock J. Reine Angew. Math. \textbf{ 658}, 193--245 (2011).

\bibitem[MSS12]{Mylonas:2012pg}
D.~Mylonas, P.~Schupp and R.~J. Szabo, \textsl{ {Membrane Sigma-Models and
  Quantization of Non-Geometric Flux Backgrounds}},
\newblock JHEP \textbf{ 1209}, 012 (2012), {1207.0926}.

\bibitem[MSS13]{Mylonas:2013jha}
D.~Mylonas, P.~Schupp and R.~J. Szabo, \textsl{ {Non-Geometric Fluxes,
  Quasi-Hopf Twist Deformations and Nonassociative Quantum Mechanics}},
\newblock (2013), {1312.1621}.

\bibitem[Roy]{Deethesis}
D.~Roytenberg, \textsl{ {Courant algebroids, derived brackets and even
  symplectic supermanifolds}},
\newblock {math/9910078}.

\bibitem[Roy02a]{Roytenberg:2002nu}
D.~Roytenberg, \textsl{ {On the structure of graded symplectic supermanifolds
  and Courant algebroids}},
\newblock (2002), {math/0203110}.

\bibitem[Roy02b]{roytenberg:structure}
D.~Roytenberg,
\newblock On the structure of graded symplectic supermanifolds and {C}ourant
  algebroids,
\newblock in \textsl{ Quantization, {P}oisson brackets and beyond
  ({M}anchester, 2001)}, volume 315 of \textsl{ Contemp. Math.}, pages
  169--185, Amer. Math. Soc., Providence, RI, 2002.

\bibitem[Roy02c]{MR1958835}
D.~Roytenberg,
\newblock On the structure of graded symplectic supermanifolds and {C}ourant
  algebroids,
\newblock in \textsl{ Quantization, {P}oisson brackets and beyond
  ({M}anchester, 2001)}, volume 315 of \textsl{ Contemp. Math.}, pages
  169--185, Amer. Math. Soc., Providence, RI, 2002.

\bibitem[Roy02d]{Roytenberg:2001am}
D.~Roytenberg, \textsl{ {Quasi Lie bialgebroids and twisted Poisson
  manifolds}},
\newblock Lett.Math.Phys. \textbf{ 61}, 123--137 (2002), {math/0112152}.

\bibitem[Sch99]{Schomerus:1999ug}
V.~Schomerus, \textsl{ {D-branes and deformation quantization}},
\newblock JHEP \textbf{ 9906}, 030 (1999), {hep-th/9903205}.

\bibitem[Sie93]{Siegel:1993th}
W.~Siegel, \textsl{ {Superspace duality in low-energy superstrings}},
\newblock Phys.Rev. \textbf{ D48}, 2826--2837 (1993), {hep-th/9305073}.

\bibitem[Ste12]{stern:short}
D.~Sternheimer,
\newblock A very short presentation of deformation quantization, some of its
  developments in the past two decades, and conjectural perspectives,
\newblock in \textsl{ Travaux math\'ematiques. {V}olume {XX}}, volume~20 of
  \textsl{ Trav. Math.}, pages 205--228, Fac. Sci. Technol. Commun. Univ.
  Luxemb., Luxembourg, 2012.

\bibitem[STW05]{Shelton:2005cf}
J.~Shelton, W.~Taylor and B.~Wecht, \textsl{ {Nongeometric flux
  compactifications}},
\newblock JHEP \textbf{ 0510}, 085 (2005), {hep-th/0508133}.

\bibitem[STW07]{Shelton:2006fd}
J.~Shelton, W.~Taylor and B.~Wecht, \textsl{ {Generalized Flux Vacua}},
\newblock JHEP \textbf{ 0702}, 095 (2007), {hep-th/0607015}.

\bibitem[SW99]{Seiberg:1999vs}
N.~Seiberg and E.~Witten, \textsl{ {String theory and noncommutative
  geometry}},
\newblock JHEP \textbf{ 9909}, 032 (1999), {hep-th/9908142}.

\bibitem[Vai12]{Vaisman:2012ke}
I.~Vaisman, \textsl{ {On the geometry of double field theory}},
\newblock J.Math.Phys. \textbf{ 53}, 033509 (2012), {1203.0836}.

\bibitem[{Vor}05a]{zbMATH02217997}
T.~{Voronov}, \textsl{ {Higher derived brackets and homotopy algebras.}},
\newblock {J. Pure Appl. Algebra} \textbf{ 202}(1-3), 133--153 (2005).

\bibitem[Vor05b]{MR2223157}
T.~T. Voronov,
\newblock Higher derived brackets for arbitrary derivations,
\newblock in \textsl{ Travaux math\'ematiques. {F}asc. {XVI}}, Trav. Math.,
  XVI, pages 163--186, Univ. Luxemb., Luxembourg, 2005.

\bibitem[Zwi12]{Zwiebach:2011rg}
B.~Zwiebach, \textsl{ {Double Field Theory, T-Duality, and Courant Brackets}},
\newblock Lect.Notes Phys. \textbf{ 851}, 265--291 (2012), {1109.1782}.

\end{thebibliography}

\end{document}